\documentclass[aps,nofootinbib,twocolumn,notitlepage,floatfix,superscriptaddress,secnumarabic]{revtex4-1}

\usepackage{setspace}
\usepackage{amsfonts,amssymb}
\usepackage{graphicx}
\usepackage[latin2]{inputenc}
\usepackage{t1enc}
\usepackage{MnSymbol}
\usepackage{verbatim}
\usepackage[    bookmarks,
                 bookmarksopen = true,
                 bookmarksnumbered = true,
                 breaklinks = true,
                 linktocpage,
                 colorlinks = true,
                 linkcolor = blue,
                 urlcolor  = blue,
                 citecolor = blue,
                 anchorcolor = green,
                 hyperindex = true,
                 hyperfigures]
		{hyperref}

\usepackage{color}

\newcommand{\dd}{\textmd{d}}
\newcommand{\be}{\begin{equation}}
\newcommand{\ee}{\end{equation}}
\newcommand{\Z}{\mathcal{Z}}

\newcommand{\expv}[1]{\left \langle #1 \right \rangle}

\begin{document}

\title{Regressive and generative neural networks for scalar field theory}

\author{Kai Zhou}
\thanks{Email: zhou@fias.uni-frankfurt.de}
\affiliation{Frankfurt Institute for Advanced Studies, 60438 Frankfurt am Main, Germany}
\affiliation{Institut f\"ur Theoretische Physik, Goethe Universit\"at, 60438 Frankfurt am Main, Germany}

\author{Gergely Endr\H{o}di}
\affiliation{Institut f\"ur Theoretische Physik, Goethe Universit\"at, 60438 Frankfurt am Main, Germany}

\author{Long-Gang Pang}
\affiliation{Frankfurt Institute for Advanced Studies, 60438 Frankfurt am Main, Germany}
\affiliation{Department of Physics, University of California, Berkeley, CA 94720, USA}
\affiliation{Nuclear Science Division, Lawrence Berkeley National Laboratory, Berkeley, CA 94720, USA}

\author{Horst St\"ocker}
\affiliation{Frankfurt Institute for Advanced Studies, 60438 Frankfurt am Main, Germany}
\affiliation{Institut f\"ur Theoretische Physik, Goethe Universit\"at, 60438 Frankfurt am Main, Germany}
\affiliation{GSI Helmholtzzentrum f\"ur Schwerionenforschung, 64291 Darmstadt, Germany}

\begin{abstract}
We explore the perspectives of machine learning techniques in the context of quantum field theories. In particular, we discuss two-dimensional complex scalar field theory at nonzero temperature and chemical potential -- a theory with a nontrivial phase diagram. A neural network is successfully trained to recognize the different phases of this system and to predict the value of various observables, based on the field configurations. We analyze a broad range of chemical potentials and find that the network is robust and able to recognize patterns far away from the point where it was trained. Aside from the regressive analysis, which belongs to supervised learning, an unsupervised generative network is proposed to produce new quantum field configurations that follow a specific distribution. An implicit local constraint fulfilled by the physical configurations was found to be automatically captured by our generative model. We elaborate on potential uses of such a generative approach for sampling outside the training region.
\end{abstract}

\maketitle

\section{Introduction}

Deep learning with a hierarchical structure of artificial neural networks is a branch of machine learning aiming at understanding and extracting
high-level representations of big data~\cite{Hinton:2015nature}. It is particularly effective in tackling complex non-linear systems with a high level of
correlations that cannot be captured easily by conventional techniques. 
Traditionally employed for tasks like pattern recognition in images or speech, automated translation or board game-playing, 
applications of 
deep learning have been found recently in many areas of physics including nuclear~\cite{Pang:2016vdc,Utama:2017ytc,Bayram:2018gqi,Neufcourt:2018syo}, particle~\cite{Baldi:2014kfa,Baldi:2014pta,Barnard:2016qma,Moult:2016cvt,Radovic:2018dip} and condensed matter~\cite{2016PhRvB..94s5105W,2017NatPh..13..435V,2017PhRvX...7c1038C,2017NatSR...7.8823B,2017NatPh..13..431C,Carleo602,2017PhRvB..95d1101L,2016PhRvB..94p5134T,2017PhRvB..95c5105H,PhysRevX.8.031012} physics. 

Significant progress has been made in utilizing machine learning methods for condensed matter systems like classical or quantum spin models. 
Specific tasks in these settings include the discrimination between certain phases and the identification of 
phase transitions~\cite{2016PhRvB..94s5105W,2017NatPh..13..435V,2017PhRvX...7c1038C,2017NatSR...7.8823B,2017NatPh..13..431C}, 
the compressed representation of quantum wave functions~\cite{Carleo602} 
or the acceleration of Monte-Carlo algorithms~\cite{2017PhRvB..95d1101L,2016PhRvB..94p5134T,2017PhRvB..95c5105H}.
Recently, deep neural networks have also been considered in particle physics, 
for the processing of experimental heavy-ion collision datasets~\cite{Pang:2016vdc}
and in the context of algorithmic development for numerical lattice field
theory simulations~\cite{Mori:2017nwj,Shanahan:2018vcv,Tanaka:2017niz}.

Pattern recognition, especially classification and regression tasks, have been discussed previously in interacting many-body
systems for condensed matter physics. In the present paper, we generalize the application of deep learning for the classification of
phases in a lattice quantum field theoretical setting. 
We further demonstrate
the capability of deep neural networks in learning physical observables, even with highly non-linear dependence on the field
configurations and with only limited training data -- 
providing an effective high-dimensional
non-linear regression method. In addition, we proceed by
implementing, for the first time, a Generative Adversarial Network (GAN)~\cite{NIPS2014_5423} for lattice field theory in order to
generate field configurations following and generalizing the training set distribution. This is an unsupervised learning framework that uses
unlabeled data to perform representation learning.
Such a GAN-powered approach is not a
full-fledged alternative to the Monte-Carlo algorithm, which possesses desired properties like ergodicity, reversibility and detailed
balance. However, it can result in a one-pass direct sampling network where no Markov chain is needed. Our aim here is to provide a proof of principle that generative networks, if trained adequately, are capable of capturing and representing
the distribution of configurations in a strongly correlated quantum field theory. On the practical side, generative networks
would prove useful when combined with traditional approaches to accelerate simulation algorithms, 
e.g., by improving decorrelation for proposals
in a Markov chain process. Further potential use of such setups would be for reducing large ensembles of field 
configurations into a single (highly trained) network as an efficient representation for the quantum statistical field ensembles, thereby 
significantly reducing storage requirements.

Specifically, we consider two-dimensional quantum scalar field theory
discretized on a lattice, 
and implement a deep neural network for the investigation of field configurations 
generated via standard Monte-Carlo algorithms.
We aim at testing whether the machine is capable of recognizing known 
features of the system including phase transitions and the corresponding 
behavior of various observables. More interestingly, we also look for 
hidden patterns discovered by the network, i.e.\ correlations of the 
gross features of the system with further low-level variables. Furthermore, we explore
the approach of generating field configurations using GAN and reproducing physical 
distributions. 

This paper is structured as follows. In Sec.~\ref{sec:sft} we outline
the scalar field theory setup, including the details of the 
configuration space and the definition of the observables. This is 
followed by Sec.~\ref{sec:sft_nn}, where we describe our neural network 
hierarchy and discuss the classification and regression tasks, together 
with the details of the generative network approach. In Sec.~\ref{sec:conclusions}
we summarize our main findings and conclude. Two appendices contain
the details of the dualization of scalar field theory (App.~\ref{app:sft})
and the specifications of the GAN approach (App.~\ref{app:GAN}). 

\section{Observables in scalar field theory}
\label{sec:sft}

We consider a complex scalar field $\phi$ with mass $m$ and quartic coupling $\lambda$ in 
$1+1$ dimensional Euclidean space-time at nonzero temperature $T$.
This system is studied in the grand canonical approach, introducing a 
chemical potential $\mu$ that controls how the charge density $n$ fluctuates. 
For low temperatures, two different regimes
of the parameter space can be distinguished: At low $\mu$ the density is suppressed, 
usually referred to as the Silver Blaze behavior~\cite{PhysRevLett.91.222001},
whereas above a threshold $\mu>\mu_{\rm th}$ the density increases considerably.\footnote{In the following we will refer to the pronounced 
change in the density at the threshold as a transition, keeping in mind that 
-- in accordance with the Mermin-Wagner theorem -- it is not connected 
to spontaneous symmetry breaking.}
People conjecture that the QCD phase diagram also holds such a behavior in the region
at low temperatures and medium to high densities~\cite{PhysRevLett.91.222001}.

This interesting behavior is a non-perturbative phenomena and cannot be observed directly, as the action becomes complex 
for $\mu\neq 0$,
hindering standard simulations in terms of the field $\phi$. 
However, using the worldline formalism, the partition function
can be reexpressed using dual variables and the 
action rendered real and positive~\cite{Gattringer:2012df}, see details 
in App.~\ref{app:sft}. The dual variables 
are the integers $k_\nu(x)$ and $\ell_\nu(x)$ that are associated to the links 
starting at the point $x=(x_1,x_2)$ and lying in the direction $\nu=1$ (space) or $\nu=2$ (time). 
The total number of variables is therefore $\mathcal{N}\equiv 2\times 2\times N_1\times N_2$, where $N_\nu$ denotes 
the number of lattice sites in the direction $\nu$. 
The partition function becomes a sum over this $\mathcal{N}$-dimensional configuration 
space,
\be
\Z = \sum_{\{k,\ell\}} \exp\left( -S^{\rm lat}[k,\ell\,] \right)\,,
\label{eq:Zdef}
\ee
with the lattice action $S^{\rm lat}$ described in Eq.~(\ref{eq:Zlatt}).
While the $\ell$-integers can take arbitrary values, the $k$-integers 
satisfy a zero divergence-type constraint,
\be
\sum_\nu [ k_\nu(x)-k_\nu(x-a\hat\nu)] = 0\,,
\label{eq:divergence}
\ee
where $\hat\nu$ is the unit vector in the $\nu$ direction and $a$ the lattice spacing. 

The partition function~(\ref{eq:Zdef}) contains all information about the system. 
In particular, the expectation values of the particle density and of the 
squared field are related to derivatives of $\Z$ with respect to $\mu$, and to $m^2$, 
respectively. Using the explicit form of the action~(\ref{eq:Zlatt}), the
corresponding operators read
\begin{align}
\label{eq:obs1}
n&= \frac{1}{N_1N_2a}\sum_x k_2(x)\,, \\
|\phi|^2&=\frac{1}{N_1N_2}\sum_x \frac{W[s(k,\ell;x)+2]}{W[s(k,\ell;x)]}\,,
\label{eq:obs2}
\end{align}
where the weight $W[s]$ and the function $s(k,\ell;x)$ are defined in 
Eq.~(\ref{eq:Ws}).

Summarizing, in our representation of complex scalar field theory, one field configuration 
consists of $\mathcal{N}$ integers $k_\nu$ and $\ell_\nu$ 
and the path integral is a sum over these 
field configurations, generated with the appropriate probabilities.
The observables on a given configuration are obtained according to Eqs.~(\ref{eq:obs1})-(\ref{eq:obs2}). This is a simple sum over the $k_2$ variables 
for the density and a highly nonlinear function for the squared 
field operator that depends on all $k_\nu$ and $\ell_\nu$ variables, see~(\ref{eq:Ws}). 

\begin{figure}[t]
	\centering
	\includegraphics[width=.9\columnwidth]{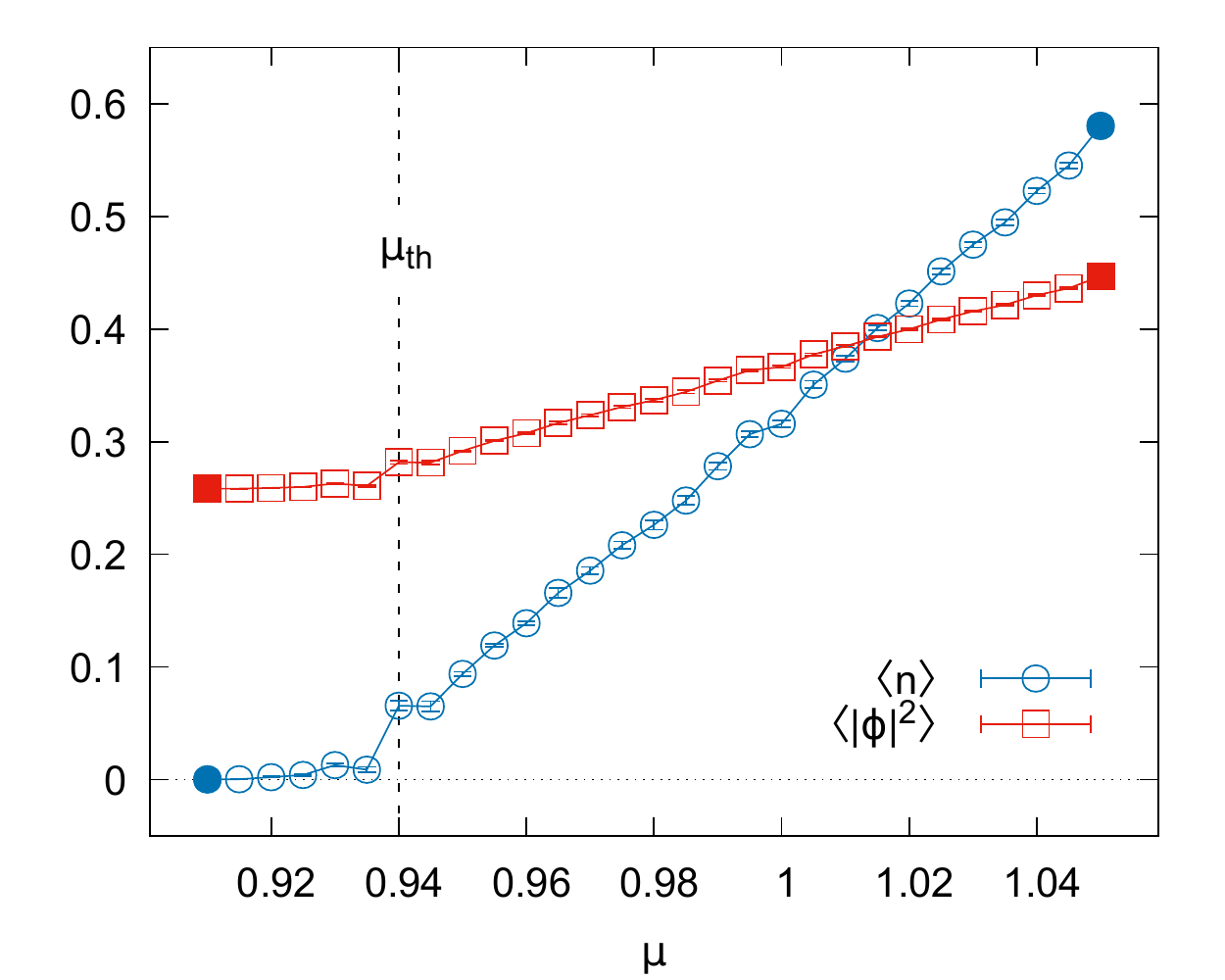}
	\caption{The expectation values of the density and of the squared field, as
		functions of the chemical potential, connected by lines to guide the eye.
		The dashed vertical line marks the threshold 
		chemical potential $\mu_{\rm th}$. The filled symbols indicate ensembles that are used in the 
		training of our neural network (see details in the text). }
	\label{fig:observables}
\end{figure}

We consider a low-temperature ensemble $N_1\times N_2=10\times 200$ 
(the dimensionality of the configuration space is therefore $\mathcal{N}=8000$)
generated with 
mass $m=0.1$, coupling $\lambda=1.0$ and a range of chemical potentials $0.91\le \mu\le 1.05$ around the threshold value $\mu_{\rm th}\approx 0.94$ (all dimensionful 
quantities are understood in lattice units). 
For $\mu<\mu_{\rm th}$, $\expv{n}$ is almost zero and $\expv{\phi^2}$ is constant. 
In contrast, both observables rise approximately linearly beyond the threshold. 
This is demonstrated in Fig.~\ref{fig:observables}.
 
\begin{figure*}
	\centering
	\vspace*{-1.cm}
	\includegraphics[trim={0 6.5cm 0 5cm},clip,width=0.85\textwidth]{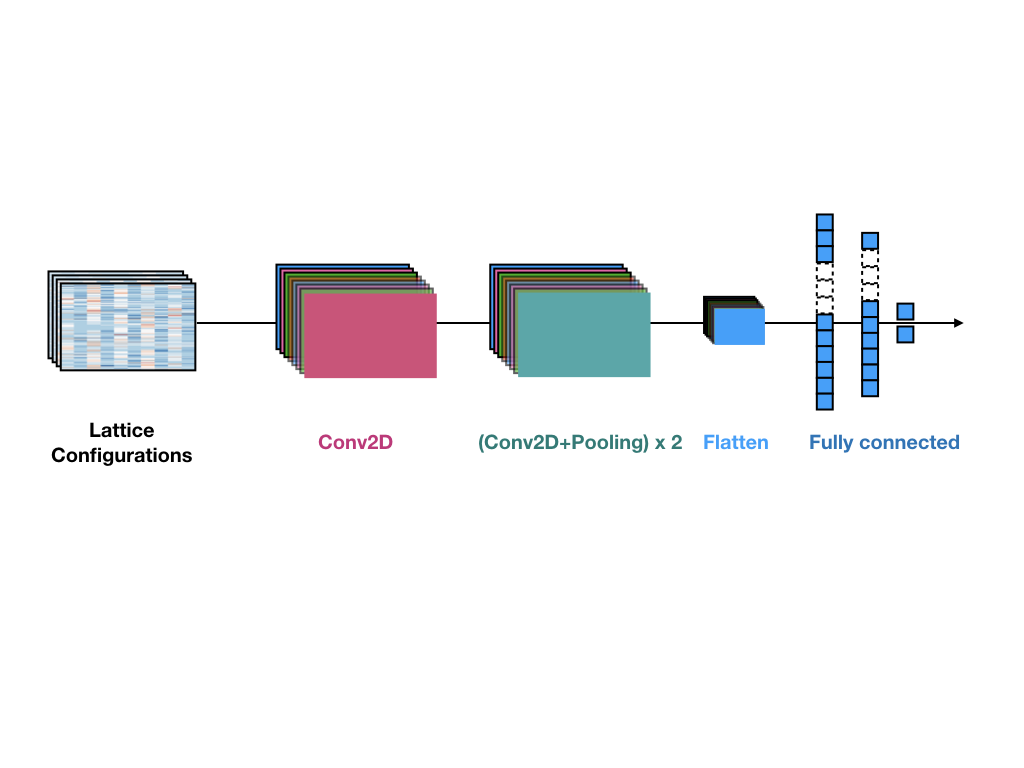}\vspace*{-1.4cm}
	\caption{The architecture of our convolutional neural network.
		The input configurations are visualized according to the lattice geometry, with the coloring reflecting
		the values of the integers sitting on the links (warm colors correspond to positive and cold colors 
		to negative values).
		The output is a binary vector describing the preference of the network, whether the configuration belongs to the low-density or 
		to the high-density phase.}
	\label{cnn}
\end{figure*}

An additional remark about the density operator~(\ref{eq:obs1}) is in order.
Due to the constraint~(\ref{eq:divergence}), the $k_\nu$ variables always
form closed loops. The contribution of such loops to $n$ depends on how many 
times the loop winds around the $\nu=2$ direction. Therefore, 
the density operator may only assume $1/N_1$ times integer values on any 
configuration. For the presently investigated lattice geometry 
this means that $n$ is an integer multiple of $0.1$. Note that this discreteness of 
the density is a finite volume artifact and $n$ becomes continuous in the 
thermodynamic limit.

\section{Scalar field theory in a neural network}
\label{sec:sft_nn}

In the present work we apply deep neural networks for the complex scalar field configurations generated with the dualization approach using
standard Monte-Carlo methods, as described in Sec.~\ref{sec:sft}. Specifically, the lattice configurations consisting of $\mathcal{N}$ integers are considered as
input to the machine. We investigated the ability of the neural network to 
perform different tasks including 
phase transition detection and physical observable regression. 
Finally we propose a new configuration generation method using the Generative Adversarial Network (GAN) approach.

\subsection{Classification of phases}
\label{sec:class}

We first employ the network to detect the transition between the low- and high-density 
phases of the system by performing a classification task. 
In particular, we train the neural network to identify the threshold chemical potential
$\mu_{\rm th}$ without specific physical guidance. 
The recognition of high-level abstract patterns in the data is essential for this classification,
thus we consider a Convolutional Neural Network (CNN), which is usually designed for 
such tasks. 
We train a CNN to target at a binary classification: The configurations are either in the low-density ``Silver Blaze'' region ($\expv{n}\approx 0$) or
in the condensed region ($\expv{n}\neq0$). To perform a semi-supervised training, we feed the lattice configurations at $\mu=0.91\ll\mu_{\rm th}$
and at $\mu=1.05\gg\mu_{\rm th}$ as input to the CNN network which has a topology as shown
in Fig.~\ref{cnn}. The training points are also highlighted in Fig.~\ref{fig:observables}.

The input configurations are viewed as images with 4 channels representing the 4 integer field variables ($k_1$, $k_2$, $\ell_1$ and $\ell_2$) and lattice size $200\times10$.
We use three convolutional layers each followed by average pooling (except for the first convolutional layer, see in Fig.~\ref{cnn}), batch normalization (BN), dropout and
PReLU activation. In the first convolutional layer there are 16 filters of size $3\times3$ scanning through the input
configuration images and creating 16 feature maps of size $200\times10$. After BN and PReLU activation, these feature maps are further convoluted in the second convolutional layer with 32 filters of size $3\times3\times32$. The output from second convolutional layer are half pooled by a subsequent average pooling layer before further processing. Dropout is applied after the final convolutional layer and in between the first two fully-connected layers. The weight matrix of both
convolutional layers are initialized with normal distribution and constrained with $L_2$ regularization. 
In a convolutional layer, each neuron only locally connects to a small chunk of neurons in the previous layer by a
convolution operation --- this is a key reason for the success of the CNN architecture.
After the third convolutional layer and a second average pooling the resulting 32 feature
maps of size $50\times2$ are flattened and connected to a 256-neuron fully connected layer with batch normalization, dropout and PReLU
activation. The final output layer is another fully connected layer with softmax activation and 2 neurons to indicate the two configuration classes. 
Dropout, batch normalization, PReLU and $L_2$ regularization work together to prevent overfitting that
may hinder the generalization ability of the network.

\begin{figure*} 
	\centering
	\includegraphics[width=\textwidth]{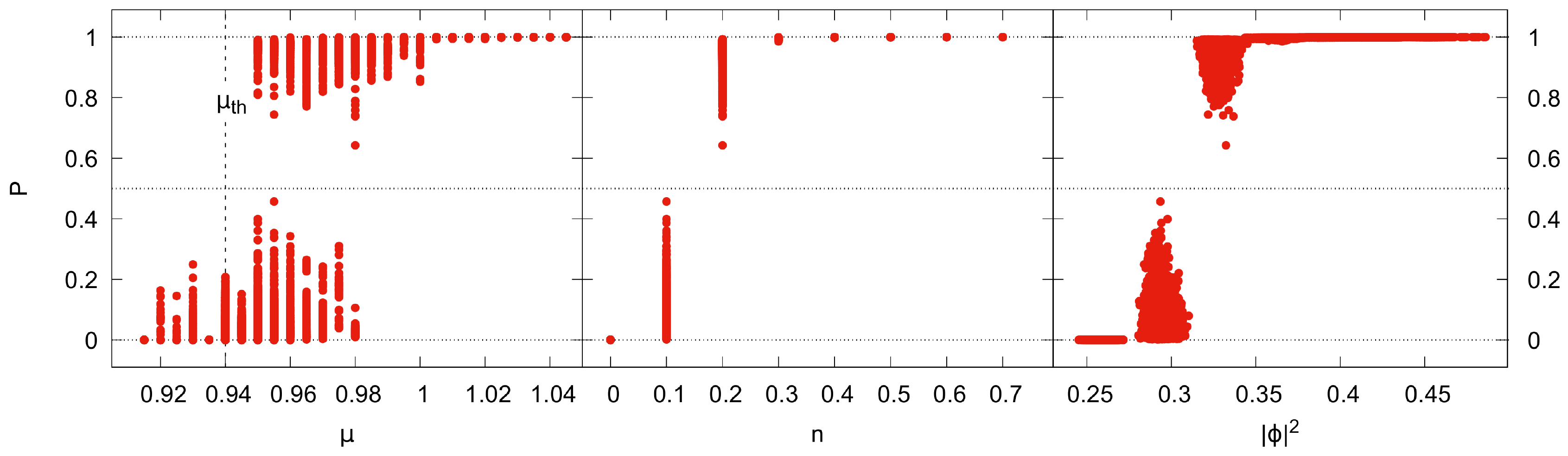} 
	\caption{The network predicted condensation probability $P$ as a function of the chemical potential (left), the particle number density (middle) and the
		squared field (right). The dashed vertical line indicates 
		the threshold chemical potential $\mu_{\rm th}$. Each point in the plot represents one configuration. As pointed out in Sec.~\ref{sec:sft}, 
		$n$ only assumes values that are integer multiples of $0.1$, 
		visible in the middle panel. }
	\label{prob2}
\end{figure*}

Supervised learning is applied here for this binary classification problem, where the configurations at $\mu=0.91$ are labelled as $(0,1)$ and the ones at $\mu=1.05$ as $(1,0)$ in the training dataset. The cross entropy between the true label and the network output (binary vector), which can
quantify well the difference for distributions, is taken as the loss function $l(\theta)$ for training the network where $\theta$ represents the trainable
parameters of the neural network. Learning/Training is performed by updating $\theta\to\theta-\alpha\partial l(\theta)/\partial\theta$ to minimize the loss function, where $\alpha$ is the learning rate with initial value 0.0001 and adaptively changed using the AdaMax method. The training data set consists of
30,000 configuration samples for each class and fed into the network in batches with batch size selected to be 16. 20\% of the training set are randomly chosen to serve as validation set. In our study, the training runs for 1000 epochs for the neural network, during which the model parameters are saved to a new checkpoint whenever a smaller validation error is encountered. Small fluctuations of validation accuracy are observed to saturate
at around 99\%.

Once trained, we test the CNN by scanning through the configurations at different values of the chemical potential $0.91<\mu<1.05$. The output of the network for each configuration 
is identified as the probability $P$ that the configuration in question corresponds to 
the condensed phase. 
In Fig.~\ref{prob2} we show $P$ predicted by the network as a function of 
various quantities: the chemical potential, the number density and the squared field. Looking at $P(\mu)$ in the left panel of the figure, we observe that while for low/high chemical potentials
the configurations unambiguously fall in one of the two classes ($P=0$/$P=1$), the ensembles 
at intermediate $\mu$ contain configurations from both sectors. This expresses 
the enhanced fluctuations in the vicinity of $\mu=\mu_{\rm th}$, as expected near 
a transition. 

To understand what the deep neural network has learned for its decision making on phase classification, we can investigate the correlation between
neural network's output and physical observables. Indeed, much more interesting trends are visible in the plots showing $P$ as a function of 
$n$ and of $|\phi|^2$ as shown in Fig.~\ref{prob2} (middle and right panels). The network outputs are strongly correlated with $n$ and $|\phi|^2$.
Without any specific supervision to the network about their role, 
the CNN has clearly learned the relevance of these observables for 
the transition. In other words, the designed CNN managed to identify highly non-linear 
features in the configurations that correspond to the physical observables~(\ref{eq:obs1})
and~(\ref{eq:obs2}). Particularly for the number density $n$, we see that its non-zero value is perfectly indicated by the non-zero
probability $P$ from the trained network.

We mention that we also tried reducing the number of convolutional layers in the network to two. 
In this case we observed similar signals for $P$ with slightly worse performance. 

\begin{figure}[b]
	\centering
	\includegraphics[width=.9\columnwidth]{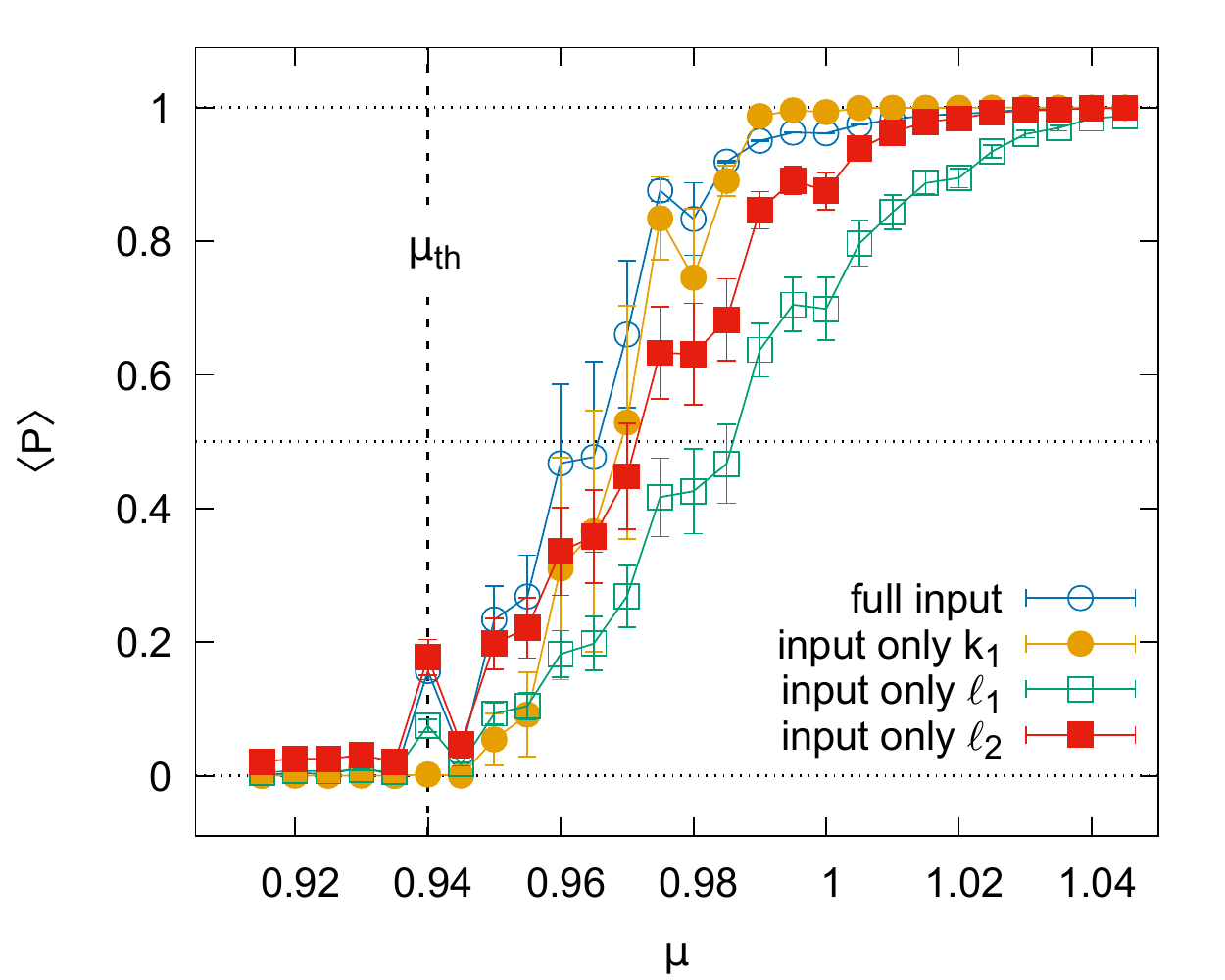} 
	\caption{The expectation value $\expv{P}$ of the condensation probability as a function of the chemical potential using full and 
		restricted inputs. The error bar shows the statistical error in one ensemble.
		The dashed vertical line marks the threshold
		chemical potential and the 
		lines connecting the points serve to guide the eye.}
	\label{prob3}
\end{figure}

\begin{figure}[t]
	\centering
	\includegraphics[width=.9\columnwidth]{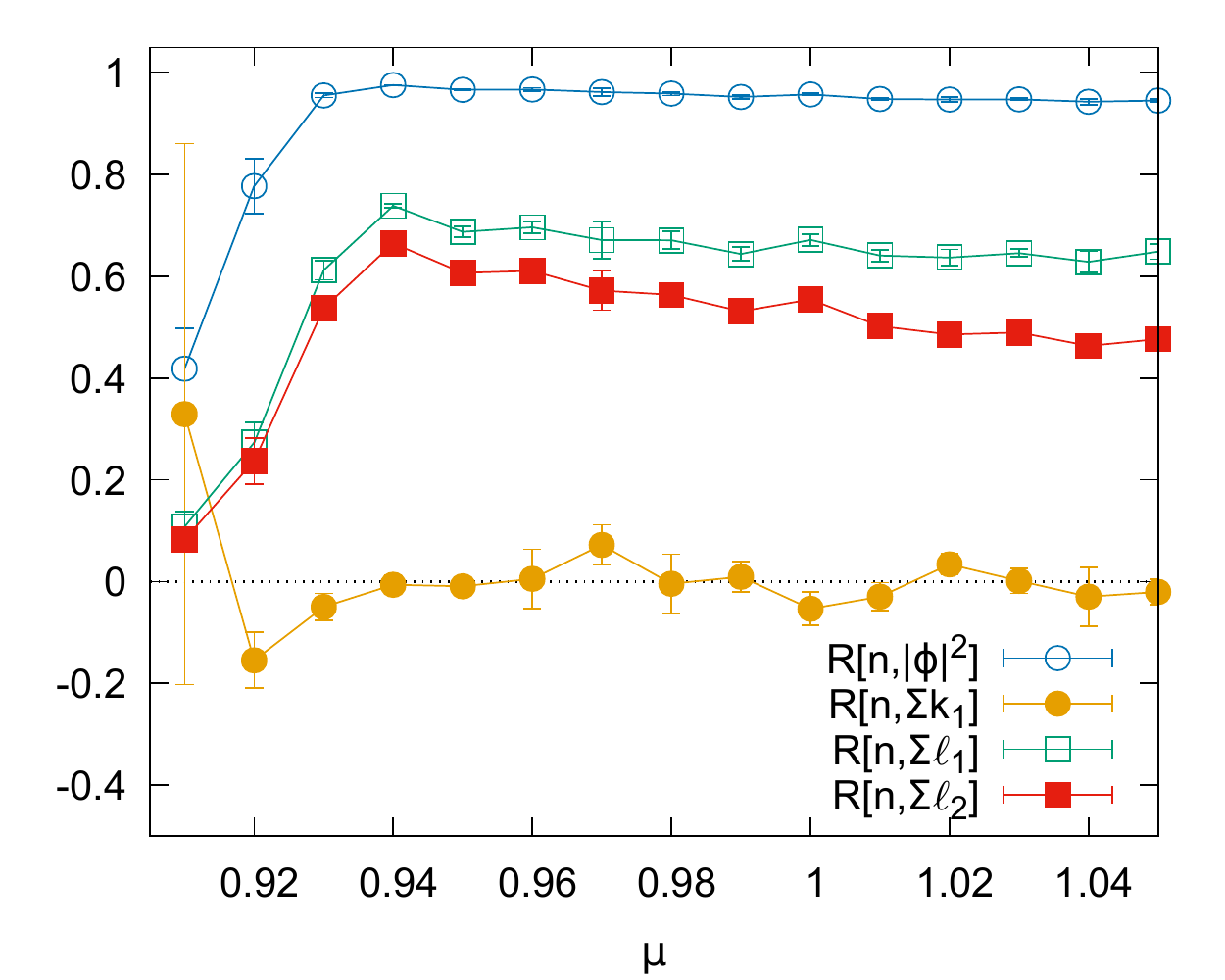} 
	\caption{The normalized correlation coefficient~(\protect\ref{eq:Rdef}) of the 
	number density and various observables including $|\phi|^2$ and the sum of the 
	$k_1$, $l_1$ or $l_2$ variables over all lattice sites.}
	\label{correlation}
\end{figure}

As can be read off from Eq.~(\ref{eq:obs1}), the particle number is given by the sum of all the $k_2$
variables. This simple pattern might be easily learned by the network. We thus performed the
same binary classification task with a restricted training input, including 
only one of the remaining three variable sets: either $k_1$, $\ell_1$ or $\ell_2$. 
The results for the three restricted inputs, together with the full 
input, are visualized in Fig.~\ref{prob3}, plotting the expectation value of the 
network predicted condensation probability against the chemical potential. Clearly, the network succeeded 
in learning essentially the same features also using the restricted inputs,
as $\expv{P}$ starts to rise at around the same threshold chemical potential $\mu_{\rm th}\approx 0.94$ for all four cases. 
This inspires us to analyze the correlation between the number density
and the similarly defined observables involving 
either $k_1$, $\ell_1$ or $\ell_2$. 
To this end we consider the normalized correlation coefficient
\be
R[A,B] \equiv \frac{\expv{AB}-\expv{A}\expv{B}}{\sqrt{\expv{A^2}-\expv{A}^2}\sqrt{\expv{B^2}-\expv{B}^2}}\,,
\label{eq:Rdef}
\ee
which vanishes for decorrelated data and equals unity for complete correlation.
As shown in Fig.~\ref{correlation},
$\sum l_1$, $\sum l_2$ and also $|\phi|^2$ are all strongly correlated with $n$, while $\sum k_1$ is fully decorrelated.
Still, the machine succeeds in classifying the configurations based only on $k_1$, as indicated in Fig.~\ref{prob3}.
Note that with conventional techniques, neither of the physical observables~(\ref{eq:obs1})-(\ref{eq:obs2}) sensitive to the transition 
can be constructed using only the $k_1$ variables. 
The excellent performance of our CNN, shown in Fig.~\ref{prob3}, indicates 
the existence of strong hidden features in the $k_1$ variables that correlate with the phase of the system. 
According to these results, the network has the ability to decode these hidden correlations in a highly effective manner.

\subsection{Non-linear regression of observables}

\begin{figure}[b]
	\centering
	\includegraphics[width=0.9\columnwidth]{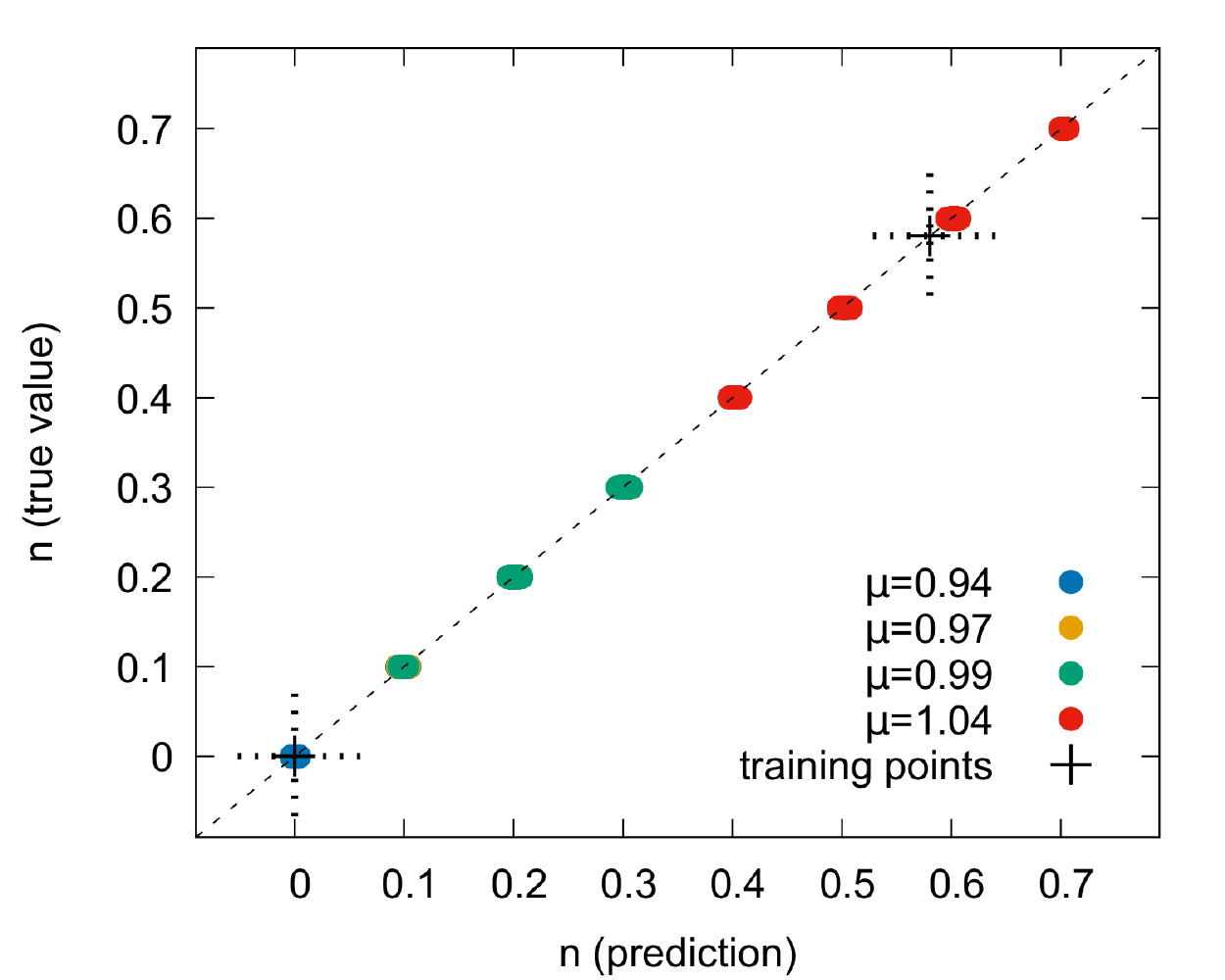}
	\hspace*{-.3cm}\includegraphics[width=0.9\columnwidth]{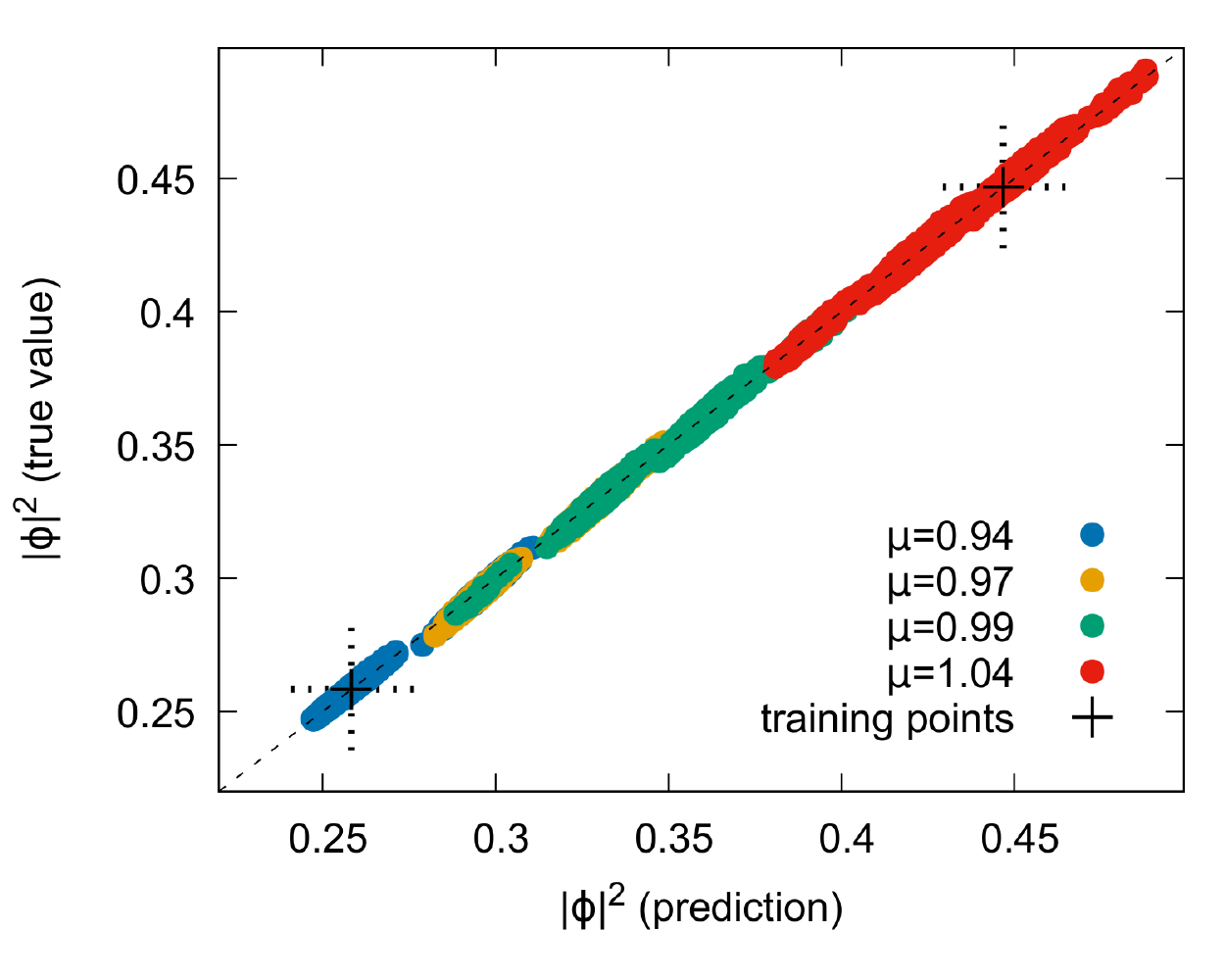}
	\caption{Comparision of the true values of the observables (vertical axis) to the network predictions (horizontal axis) for the particle density $n$ (upper panel) and for the squared field $|\phi|^2$ (lower panel). 
		Each filled circle represents one configuration and 
		the dotted crosses indicate the expectation values 
		of the observables on the training ensembles.}
	\label{regress}
\end{figure}

Next, we consider a regression task for learning thermodynamic observables of the system, employing 
a similar training strategy as used above for the 
binary classification. In particular, 
supervised learning is applied
with a CNN to regress the thermodynamic observables including the particle number density and the squared field, based on the lattice configurations. 
As in Sec.~\ref{sec:class}, the training dataset consists of configurations at $\mu=0.91$ and $\mu=1.05$. The generalization ability of the machine is
investigated by testing the network predictions on configurations 
at intermediate values of the chemical potential. 

To target this regression task, we change the CNN architecture slightly. 
Specifically, the batch normalization and pooling layers are removed, and one
more fully connected layer with 32 neurons is inserted before the final output layer. The latter consists of 2 neurons representing the values of
$n$ and of $|\phi|^2$. The activation functions are all changed to ReLU and the loss function for the network is chosen
to be the mean squared difference between the predictions and the true values. After 2000 epochs of training, our regression network is tested on previously
unseen configurations at different values of the chemical potential.
The results for the density and for the squared field are plotted in Fig.~\ref{regress}, showing the true values of the
observables, calculated using Eqs.~(\ref{eq:obs1})-(\ref{eq:obs2}), against the network predictions. 

As visible in Fig.~\ref{regress}, the network performs well
over a broad range of chemical potentials,
predicting $n$ and $|\phi|^2$ accurately, the maximal deviation being 
around $5\%$ for the density and around $7\%$ for the squared field. 
Note that the training was performed using only two far-away segments of 
the range of the target observables, corresponding to configurations
at $\mu=0.91$ and $\mu=1.05$ (the expectation values of the observables
for these ensembles are also indicated in the plots). 
On the one hand, the high quality of the regression for the density 
may seem natural owing to the linear dependence~(\ref{eq:obs1}) 
of $n$ on the individual variables. On the other hand, the squared 
field is a highly non-linear function of the high-dimensional input ($\mathbb{R}^{200\times 10\times 4}\to\mathbb{R}^{1}$), making the 
excellent predictive ability of the network very non-trivial and 
surprising.  
Put differently, 
using limited training data (covered small range of the target domain), 
our CNN network has the ability 
to correctly reproduce the whole target space mapping, which is curved and even dramatically
changing (close to transition point). This means that the network has effectively encoded the configuration into a much plainer and abstract latent space (intermediate
layers inside the network). A linear interpolation in these layers
can result in non-linear regression in the final output layer.

Just as the middle panel of Fig.~\ref{prob3}, the upper panel of Fig.~\ref{regress} reflects the discreteness of the density operator $n$, 
evaluated on any configuration. Notice that while the true values of $n$
are indeed integer multiples of $0.1$, small deviations (below $0.007$ in 
magnitude) from this rule occur for the predicted values. Such 
deviations stem from the approximative nature of the regression network and 
are observed to decrease as the number of training epochs is increased. 
We get back to this behavior below in the generative network analysis.

\subsection{Configuration production using the Generative Adversarial Network}

\begin{figure}[t]
	\centering
	\includegraphics[width=.9\columnwidth]{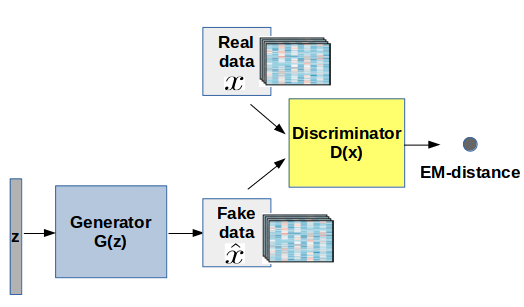}
	\caption{Architecture of a generative adversarial network for complex scalar field theory.}
	\label{fig:wgan}
\end{figure}

Generative Adversarial Network (GAN)~\cite{NIPS2014_5423} is a deep generative model that aims to learn the distribution of input variables from the training data. 
It belongs to the unsupervised learning category within deep learning approaches. The GAN framework contains two non-linear differentiable functions, both of which are represented
by adaptive deep neural networks. The first one is the generator $G(z)$, which maps random noise vectors $z$ from a latent space with distribution $p_{\rm prior}$ (usually uniform or normal distribution over $z$) to the target data space with implicit distribution $p_G$ (over data $x$) that 
approaches the desired distribution $p_{\rm true}$ through training. 
The second one is the discriminator $D(x)$ with a single scalar output, which tries to distinguish real data $x$ from generated data $\hat{x}=G(z)$.
These two neural networks are trained alternately, thus improving their respective abilities against each other in a two-player minimax game (also called zero-sum game). An optimally trained GAN converges to the state (the Nash equilibrium for this game-theory problem), 
where the generator excels in `forging' samples that the discriminator cannot 
anymore distinguish from real data.
Such generative modeling-assisted approaches have been tested in various scientific contexts, including medicine~\cite{DBLP:journals/corr/NieTPRS16,Mahapatra2017ImageSR}, particle physics~\cite{deOliveira:2017pjk,Paganini:2017dwg,Paganini:2017hrr}, cosmology~\cite{Ravanbakhsh:2016xpe,2017MNRAS.467L.110S,2017arXiv170602390M} and
condensed matter physics~\cite{PhysRevE.96.043309,PhysRevE.97.032119}.
Here we employ, for the first time, the generative modeling GAN application in strongly correlated quantum field theory.
To ensure training stability,
we consider the Wasserstein-GAN architecture~\cite{2017arXiv170107875A} with gradient penalty (WGAN-gp)~\cite{2017arXiv170400028G} in this study, see Fig.~\ref{fig:wgan} for the main architecture. The theoretical foundations of WGAN are outlined in App.~\ref{app:GAN}. 
\begin{figure}[t]
	\centering
	\includegraphics[width=.9\columnwidth]{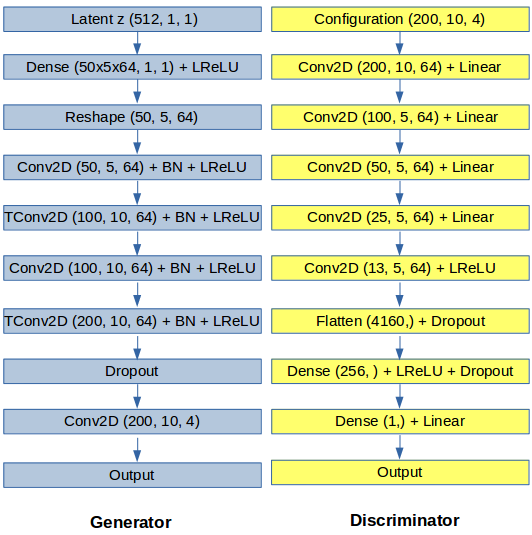}
	\caption{Illustration of our Generator and Discriminator network architectures.
		The transposed convolutional layer for upsampling is denoted as TConv2D,
		batch normalization as BN and a fully connected layer as Dense.
		For each layer, the dimensionality of the output tensor is specified in brackets.
		}
	\label{fig:GDplot}
\end{figure}

The generator and discriminator architectures are illustrated in Fig.~\ref{fig:GDplot}. The generator takes as input a randomly sampled 
512-dimensional latent vector $z$ following a multivariate normal Gaussian distribution, and gradually transforms $z$ to the desired configuration space
(of dimensionality $200\times 10\times 4$). The up-sampling is done via transposed convolution, which is
also known as fractionally-strided convolution that function backward
the convolution operation. The kernel size for the convolutional layer
is $3\times 3$, while for the transposed convolutional layer $4\times 4$. Batch normalization is included to standardize the outputs
and to stabilize training. Apart from the last layer we use the Leaky Rectified Linear Unit (LReLU) as activation function.
The discriminator
aims to evaluate the `fidelity' of the configurations. The difference between the output of real data and fake data is quantified using
the Earth Mover (EM)-distance (also called Wasserstein distance), 
which serves
as the loss function here. Strided convolution is performed for the down-sampling. Note
that for the first four convolutional layers plain linear activation is used to let the discriminator more effectively reduce the
dimensionality of the input configurations (function like PCA). This helps the GAN to capture the implicit multimodal
distribution (of physical observables), as we will see below.

\begin{figure}[b]
	\centering
	\includegraphics[width=.9\columnwidth]{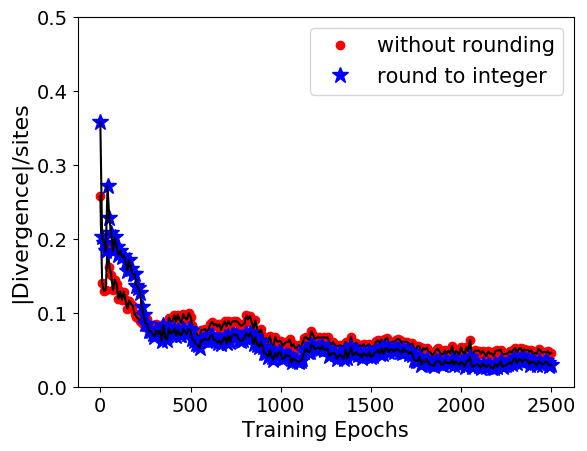}
	\caption{The absolute divergence (using LHS of Eq.(\ref{eq:divergence})) per site for configurations from the GAN generator as a function of
		training epochs, with (blue) and without (red) rounding configuration entries
		to its nearest discrete value.}
	\label{fig:div_check}
\end{figure}

Having trained the GAN, the generator can be used to convert samples from the prior distribution $p_{\rm prior}$ to data points lying in configuration space. 
To verify the effectiveness of the network and, in particular, whether the generated 
configurations are indeed physical,
we will first check the divergence-type constraint~(\ref{eq:divergence}) for the complex scalar field. As shown in Fig.~\ref{fig:div_check}, the absolute divergence per site for the generated outputs is not exactly zero but is decreasing and converges to zero as the number of training epochs grows. Note that Eq. (\ref{eq:divergence}) represents a highly implicit physical constraint inside the training dataset, which is not provided as supervision to the training of the GAN. 
Instead, the network automatically recognized this constraint for the configurations
in a converging way. The generation time for a single configuration using the GAN (on an Nvidia TitanXp GPU) is $0.2\textmd{ ms}$. 

\begin{figure}[t]
	\centering
	\includegraphics[width=.9\columnwidth]{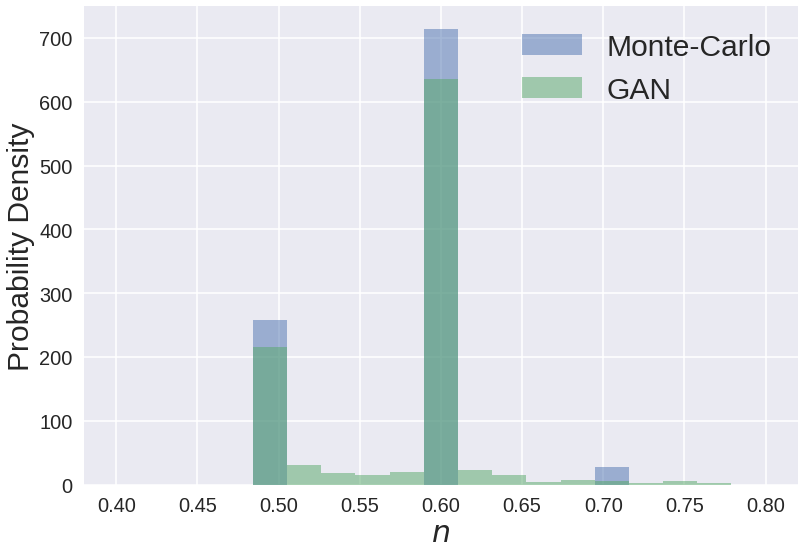}
	\includegraphics[width=.9\columnwidth]{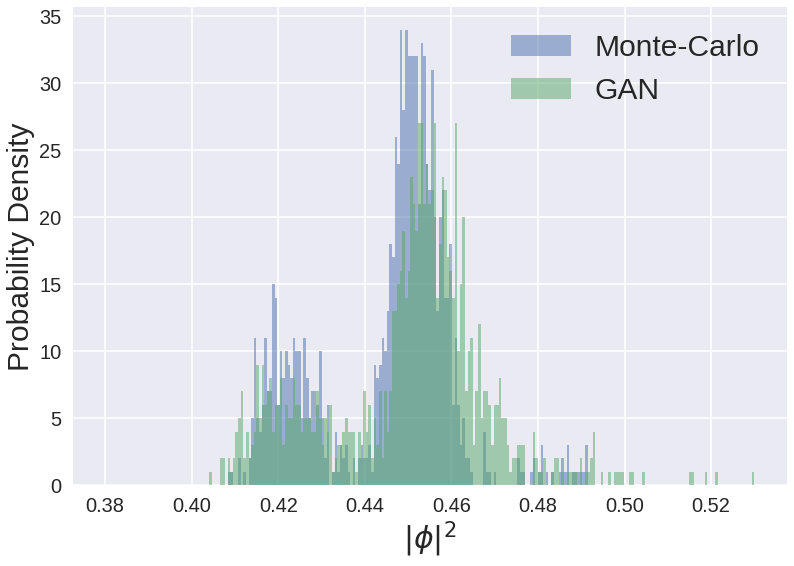}
	\caption{The probability density distribution of the number density $n$ (top panel) 
		and of the squared field $|\phi|^2$ (bottom panel) from the GAN (green) along
		with training data distribution obtained from the Monte-Carlo simulation (blue) for fixed chemical potential $\mu=1.05$
		with 1000 samples. }
	\label{fig:n_dist}
\end{figure} 

Next we turn to the distribution of observables in the samples generated by the GAN and check to 
what extent it agrees with the training distribution. In the top panel of Fig.~\ref{fig:n_dist} we visualize the probability density distribution of the number density $n$ from the GAN after training with one ensemble of configurations at $\mu=1.05$. We observe that the GAN has captured the discrete distribution of $n$ quite well. The ensemble average of the particle number density from GAN is estimated (using 1000 random samples) to be $\expv{n}_{\mathrm{GAN}}=0.578$, also quite close to the Monte-Carlo value $\expv{n}_{\mathrm{MC}}=0.580$. As mentioned earlier, the particle number density~(\ref{eq:obs1}) is simply the sum of time component of the $k$ variables in the configuration.
In contrast, the squared field $|\phi|^2$ calculated using Eq.~(\ref{eq:obs2}) is highly non-linear in 
the input variables. 
Nevertheless, the multi-modal distribution of $|\phi|^2$ is also well reproduced by the 
generative network, see the bottom panel of Fig.~\ref{fig:n_dist}. The ensemble average of $|\phi|^2$ from GAN (for the same 1000 samples above) $\langle\phi^2\rangle_{\mathrm{GAN}}=0.449$, is also close to the Monte-Carlo result $\langle\phi^2\rangle_{\mathrm{MC}}=0.447$. Figs.~\ref{fig:div_check} and~\ref{fig:n_dist} clearly demonstrate that the generative adversarial network can be trained to capture the statistical distribution of the field configurations even on the level of physical observables.

\begin{figure}[t]
	\centering
	\includegraphics[width=.9\columnwidth]{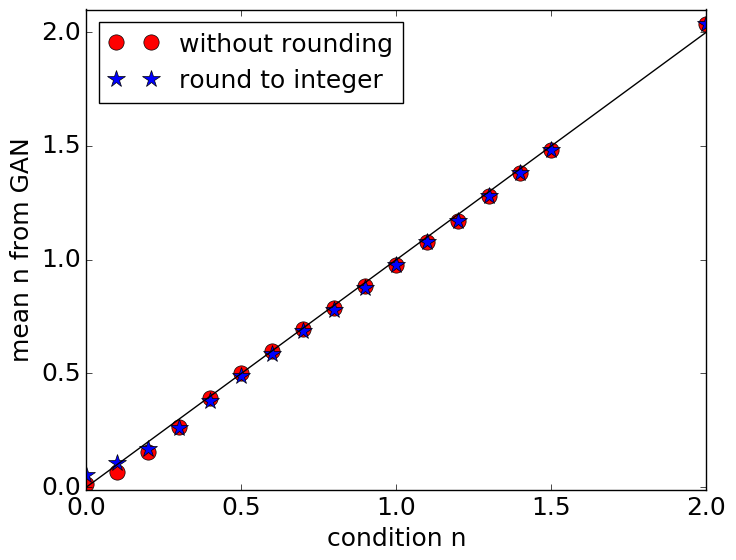}
	\caption{The mean particle number density on the configurations 
	generated by the cGAN with (blue) and without (red) rounding configuration entries
	to its nearest discrete value, against the specified condition values for $n$. }
	\label{fig:cgan_mean_n}
\end{figure}

The above GAN structure is designed to reproduce certain distributions in the training dataset. Next we attempt to use the network to generalize the distribution that it was trained on. The discriminator is provided with relevant labels (in this case the value of the number density $n$) for the training dataset in order to condition the network (cGAN)~\cite{DBLP:journals/corr/MirzaO14}. Specifically, the training sample at $\mu=1.05$  contains cases $n=0.4$, $0.5$, $0.6$ and $0.7$. After the training we test the generalization ability
of the network by specifying desired number densities outside the above set of values. Fig.~\ref{fig:cgan_mean_n} shows the performance of the cGAN for this generalization task. We stress that for the training only $0.4\le n\le0.7$ values were provided,
but the agreement between the desired $n$ (condition) and the measured $n$ on the generated 
configurations is spectacular over a much broader range of density values.
This generalization task might be viewed as converting the grand-canonical ensemble of configurations
(at fixed $\mu$) to a series of canonical ensembles (at various values of $n$).

\section{Conclusions}
\label{sec:conclusions}

In this paper we proposed a set of novel techniques for the investigation of a lattice-regularized 
quantum field theory by means of deep neural networks, including discovering hidden correlations, learning
observables and producing field configurations. 
Specifically, our analysis was carried out for the dualized representation of 
complex scalar field theory in 1+1 dimensions.

We first showed that a convolutional neural network can be used in a semi-supervised
manner to detect the phase transition in this strongly correlated quantum field theory based on 
the microscopic configurations. We found that the network
is capable of recognizing correlations in the system between various observables and phases classification without the specific knowledge guidance. Very interestingly, the network discovered
a correlation beyond the conventional analysis, which enabled it to use a restricted subset of the 
input variables (in particular, the $k_1$ variables) alone
to decode information about the phase transition.

We continued by designing a regressive neural network to learn physical observables ($n$ and $|\phi|^2$) with limited training samples. The network achieved remarkable agreement with the physical observables 
and also revealed a great generalization ability when tested at chemical potentials beyond the training
set. This approach provides an
effective high-dimensional non-linear regression method even with limited (compared to the huge Hilbert space, i.e.\ number of possible configurations) data points, where traditional interpolation or regression would require much higher statistics that grows exponentially with input dimensionality.

Finally, we proposed to generate new configurations following a specific distribution by adapting the modern deep generative modeling technique GAN. We found that
the generator in the GAN has the ability of automatically recognizing the implicit but crucial physical constraint on the configurations in an unsupervised manner,
and can represent the distribution of prominent observables with direct sampling.   
The generalization of configuration production to different parameter domains, e.g.\ towards 
a critical region, where conventional techniques slow down considerably, is clearly a fascinating 
feature that deserves further investigations.

\noindent

\acknowledgments
This work is supported by the funding of AI research at FIAS from SAMSON AG Frankfurt (K.Z.), DFG Emmy Noether Programme EN 1064/2-1 (G.E.),
NSF JETSCAPE ACI-1550228 (L.-G.P.), endowment through the Judah M. Eisenberg Laureatus Chair
at Goethe University and the Walter Greiner Gesellschaft at Frankfurt (H.St.). K.Z. gratefully acknowledges supports from the NVIDIA
Corporation with the donation of NVIDIA TITAN Xp GPU cards for the research.
G.E. acknowledges inspiring discussions with G\'abor Varga.
 
\appendix

\section{Dual formulation of scalar field theory}
\label{app:sft}

In the continuum, the Euclidean action of a complex 1+1-dimensional scalar field $\phi$ reads
\be
\int_0^L\!\! \dd x_1 \int_0^{1/T}\!\! \dd x_2 
\left[ (D_\nu \phi)^*\! (D_\nu \phi) + m^2 \phi^*\!\phi + \lambda (\phi^*\!\phi)^2\right]\,,
\label{eq:Lcont}
\ee
where the covariant derivative is $D_\nu=\partial_\nu+i\mu \delta_{\nu,2}$, $m$ denotes the mass of the charged scalar field, $\lambda$ the quartic coupling and $\mu$ the chemical potential. The spatial ($\nu=1$) and time-like ($\nu=2$) 
coordinates are labelled by $x_\nu$. In Eq.~(\ref{eq:Lcont}), $L$ denotes the spatial extent of the system and $T$ the temperature and we use periodic boundary conditions in both directions. 

On a lattice with spacing $a$, the derivative operator is discretized by nearest-neighbor hoppings. The chemical potential assigns different weights to forward and backward hoppings in the time-like direction so that the 
regularization of the continuum action is
\be
\begin{split}
S^{\rm lat} &= \sum_{x} \bigg\{ (4+m^2) \phi^*\!(x) \phi(x) + \lambda[\phi^*\!(x)\phi(x)]^2 \\
&- \sum_{\nu=1,2} \left[ e^{\mu\delta_{\nu,2}} \phi^*\!(x) \phi(x+\hat \nu) + e^{-\mu\delta_{\nu,2}} \phi^*\!(x) \phi(x-\hat\nu)\right]\bigg\}\,,
\end{split}
\ee
where $x=(x_1,x_2)$ labels the lattice sites that range over $0\le x_\nu<N_\nu$. The lattice sizes are related to the volume and the temperature as $L=N_1a$ and $T=(N_2a)^{-1}$. 
The partition function of this system is defined by the path integral~(\ref{eq:Zdef}) 
over the (complex) 
field configurations.

The chemical potential spoils the reality of $S^{\rm lat}$
so that for $\mu\neq0$ we face a complex action problem that leads to a highly oscillatory 
integrand under the path integral. This problem can be solved by the so-called worldline formalism or dualization approach~\cite{Gattringer:2012df}. 
The chief steps are an expansion of the exponential factors (for each $x$ and each $\nu$), 
a variable substitution to the polar representation $\phi=r\,e^{i\varphi}$
and a subsequent integration in $r$ and in $\varphi$. 
The final result is expressed as a sum over the integer 
expansion variables $k_\nu(x)$ and $\ell_\nu(x)$ 
and reads~\cite{Gattringer:2012df}\begin{subequations}
\label{eq:Zlatt}
\be
\Z \!=\!\! \sum_{\{k,\ell\}} \!\exp\!\left(-S^{\rm lat}[k,\ell]\right)
= \sum_{\{k,\ell\}}
\prod_x \Z^{k,\ell}(x)
\,,
\label{eq:Zlatt1}
\ee
with $\Z^{k,\ell}(x)$ given by
\be
e^{\mu k_t(x)}\cdot
W[s(k,\ell;x)] \cdot\delta[\nabla\cdot k(x)] 
\cdot\prod_\nu A[k_\nu(x),\ell_\nu(x)]\,.
\label{eq:Zlatt2}
\ee
\end{subequations}
In the second factor, $W$ is a positive weight
\be
\begin{split}
W[s] &= \!\int_0^\infty\!\!\dd r \,r^{s+1}\,e^{-(4+m^2)^2-\lambda r^4}, \\[.1cm]
s(k,\ell;x) \!&=\! \sum_\nu \left[ |k_\nu(x)|+|k_\nu(x-\hat\nu)|+2(\ell_\nu(x)+\ell_\nu(x-\hat\nu) \right]\,,
\end{split}
\label{eq:Ws}
\ee
that depends explicitly on $m$ and on $\lambda$. 
The third factor in 
Eq.~(\ref{eq:Zlatt2}) 
includes a Kronecker-$\delta$
with argument
\be
\nabla\cdot k(x) = \sum_\nu [ k_\nu(x)-k_\nu(x-\hat\nu)]\,,
\ee
which places a constraint on the $k$-variables so that their discretized divergence must vanish
at each point, as indicated in Eq.~(\ref{eq:divergence}).  
Finally, the last factor in~(\ref{eq:Zlatt2}) is a positive combinatorial factor
\be
A[k_\nu(x),\ell_\nu(x)] = \frac{1}{(\ell_\nu(x)+|k_\nu(x)|)!\;\ell_\nu(x)!}\,.
\ee

Instead of the original complex field $\phi$, the field variables are now the integers $k$ and $\ell$. 
According to Eq.~(\ref{eq:Zlatt}), the weight of any field configuration is real and positive, 
thus the system can be simulated using standard Monte-Carlo algorithms. 
Specifically, we employ a worm algorithm~\cite{Gattringer:2012df,Orasch:2017niz,Christof}, which 
is capable of automatically satisfying
the divergence constraint $\nabla\cdot k=0$ on all lattice sites. 
Differentiating the partition function~(\ref{eq:Zlatt}) we obtain the representations~(\ref{eq:obs1})-(\ref{eq:obs2}) of the
operators in terms of the integers $k$ and $\ell$.

\section{Generative Adversarial Networks}
\label{app:GAN}

The definition of the GAN involves the loss functions $\mathcal{L}_D$ and $\mathcal{L}_G$ for the discriminator and the generator, respectively. In a zero-sum game $\mathcal{L}_G=-\mathcal{L}_D$,
and upon optimization of the respective parameters $\theta_G$ and $\theta_D$ 
the game converges to
\be
\theta_{G,D}^{*}=\arg \min\limits_{\theta_G} \max\limits_{\theta_D}(-\mathcal{L}_D(\theta_G,\theta_D))\,.
\label{nash-eq}
\ee
The original GAN uses the loss function
\be
\mathcal{L}_D=-\mathbb{E}_{x\sim p_{\rm true}}[\log D(x)] - \mathbb{E}_{z\sim p_{\rm prior}}[\log(1-D(G(z)))]\,,
\label{gan_l1}
\ee
where
\be
\mathbb{E}_{x\sim p}[A]=\int \!\dd x \,p(x) A(x), \quad \int\!\dd x\,p(x)=1\,,
\ee
denotes the expectation value over the normalized probability distribution $p(x)$
and we used
\be
\mathbb{E}_{z\sim p_{\rm prior}}[A(G(z))] = \mathbb{E}_{\hat x \sim p_G} [A(\hat x)]\,.
\ee
Note that for the generator, the first term of~(\ref{gan_l1}) has no impact 
on $G$ during gradient descent updates as it only depends on $\theta_D$. The expectation values in the above loss functions are computed from the mean of all the training samples. The parameters of the discriminator and the generator are updated by back
propagation with the gradients of the loss functions,
\be
\sum_{i}\nabla_{\theta_D}\mathcal{L}_{D}(x_i,z_i;\theta_D), \quad\quad\quad \sum_{i}\nabla_{\theta_G}\mathcal{L}_{G}(z_i;\theta_G)\,,
\label{gan_grad}
\ee
where $x_i$ is the sample from the training data and $z_i$ the latent noise from the prior $p_{\rm prior}$. 
The optimal discriminator under the above well-defined loss function for given fixed generator $G$ can be derived to be
\be
D^{*}_{G}(x)=\frac{p_{\rm true}(x)}{p_{\rm true}(x)+p_G(x)}\,,
\label{dis}
\ee

From the information theory point of view, the above objective from discriminator (thus the training criterion of generator)
are nothing else but the Jensen-Shannon (JS) divergence, which is a measure of similarity between two probability distributions,
\be
\begin{split}
\max\limits_{\theta_D}(-\mathcal{L}_D(\theta_G,\theta_D))&=-\mathcal{L}_{D^{*}} \\
&= 2\,\mathbb{D}_{\rm JS}(p_{\rm true}\parallel p_G) - 2\log2\,,
\label{js}
\end{split}
\ee
and the JS divergence is formulated by symmetrizing the Kullback-Leibler (KL) divergence,
\be
\mathbb{D}_{\rm JS}(p\parallel q)=\frac{1}{2}\left[\mathbb{D}_{\rm KL}(p\parallel\frac{p+q}{2}) + \mathbb{D}_{\rm KL}(q\parallel\frac{p+q}{2})\right]\,,
\label{js2}
\ee
with the KL divergence given as
\be
\mathbb{D}_{\rm KL}(p\parallel q)=\int \!\dd x\, p(x)\log\frac{p(x)}{q(x)}\,.
\label{kl}
\ee
So the best state for the generator is reached if and only if $p_G(x)=p_{\rm true}(x)$, giving $D^{*}=1/2$ for the global optimum
of the minimax game also called Nash-equilibrium.

In practice this default setup is difficult to train especially when
applied for high-dimensional case. Using the above JS divergence measure,
the discriminator $D$ might not provide enough information to estimate the distance between the generated
distribution and the real data distribution when the two distributions do not overlap sufficiently.
Specifically, when the support of $p_G$ and $p_{\rm true}$ both rest in low dimensional manifolds of the
data space, the two distributions thus has a zero measure overlap which results in vanishing gradient for the generator. This leads to a weak signal for $G$ updating and general instability. 
Mode-collapse can easily occur for GAN where the generator learns to only produce a single element in the state space
that is maximally confusing the discriminator. In order to avoid this kind of failure training, a multitude of different techniques have been developed recently, like ACGAN~\cite{2016arXiv161009585O}, WGAN~\cite{2017arXiv170107875A},
improved WGAN~\cite{2017arXiv170400028G}, which help stabilizing and improving the GAN training. We used the improved WGAN with gradient penalty~\cite{2017arXiv170400028G} in this work. The most important difference
of WGAN compared to the original GAN lies in the loss function, where the Wasserstein-distance (also called Earth Mover distance) provides an efficient measure
for the distance between the two distributions ($p_{\rm true}$ and $p_G$) even if they are not overlapping anywhere. The loss functions are now
\be
\begin{split}
\mathcal{L}_D=&-\mathbb{E}_{x\sim p_{\rm true}}[D(x)] + \mathbb{E}_{z\sim p_{\rm prior}}[D(G(z))] \\
&+ \lambda\,\mathbb{E}_{x\sim p_{\rm true}}[\lVert \nabla_{x} D(x)
 \rVert_{p} - K]^{2}\,,
\end{split}
\label{gan_l11}
\ee
and
\be
\mathcal{L}_G= - \mathbb{E}_{z\sim p_{\rm prior}}[D(G(z))]\,,
\label{gan_l22}
\ee
where the gradient penalty term with strength $\lambda$ is computed in a linearly interpolated sample space,
\be
\hat{x}_{gp}=\epsilon x + (1-\epsilon \hat{x})\,,
\label{gp}
\ee
with uniformly sampled $\epsilon\sim(0,1]$.

\bibliographystyle{utphys}

\end{document}